\documentclass[12pt]{article}
\usepackage{fullpage,graphicx,psfrag,amsmath,amsfonts,verbatim}
\usepackage[small,bf]{caption}
\usepackage{booktabs}
\usepackage{graphicx}
\usepackage{subfigure}
\usepackage{hyperref}
\usepackage[misc]{ifsym}

\newcommand{\reals}{{\mbox{\bf R}}}

\newcommand{\eg}{{\it e.g.}}
\newcommand{\ie}{{\it i.e.}}
\newcommand{\etc}{{\it etc.}}

\newcommand{\BEAS}{\begin{eqnarray*}}
\newcommand{\EEAS}{\end{eqnarray*}}
\newcommand{\BEA}{\begin{eqnarray}}
\newcommand{\EEA}{\end{eqnarray}}
\newcommand{\BEQ}{\begin{equation}}
\newcommand{\EEQ}{\end{equation}}
\newcommand{\BIT}{\begin{itemize}}
\newcommand{\EIT}{\end{itemize}}

\title{A Markowitz Approach to Managing a Dynamic Basket of Moving-Band Statistical Arbitrages}
\author{Kasper Johansson\footnote{Stanford University (\Letter\, kasperjo@stanford.edu)} 
\and Thomas Schmelzer\footnote{Abu Dhabi Investment Authority} 
\and Stephen Boyd\footnote{Stanford University}}

\begin{document}
\maketitle

\begin{abstract}
We consider the problem of managing a portfolio of moving-band statistical
arbitrages (MBSAs), inspired by the Markowitz optimization framework. 
We show how to manage a dynamic basket of MBSAs, and illustrate
the method on recent historical data, showing that it can perform very
well in terms of risk-adjusted return, essentially uncorrelated with the market.
\end{abstract}

\newpage
\tableofcontents
\newpage

\section{Introduction}\label{s-intro} 
We consider the problem of managing a portfolio of statistical arbitrages
(stat-arbs), where each stat-arb is a mean-reverting portfolio of a subset of an
$n$-asset universe. Trading strategies based on stat-arbs have been around since the 1980s and are
popular due to their documented success in practice. The strategy is based on
the idea that if the price of a
portfolio of assets is mean-reverting, we can profit by buying the portfolio
when it is cheap and selling it when it is expensive. The literature on stat-arbs
is vast, and typically focuses on one of two aspects: (i) finding a linear
combination of asset prices that is mean-reverting, or (ii) finding a trading
policy that profits by exploiting the mean-reversion. To the best of our
knowledge, there exists no satisfying solution to the problem of managing a
portfolio of multiple stat-arbs, that is independent of the process of finding them. This paper aims to fill this gap and proposes a
solution based on a Markowitz optimization approach. In particular, we focus on
the case of moving-band stat-arbs (MBSAs), recently introduced
in~\cite{johansson2023statarb}, where the midpoint of the price band is allowed
to move over time. 

\subsection{Related work}
\paragraph{Stat-arbs.}
Stat-arb strategies have been popular ever since their introduction by Nunzio
Tartaglia in the 1980s~\cite{pole2011statistical, gatev2006pairs}. The first
strategy was based on pairs trading, where the price difference between two 
assets is tracked and positions entered when this difference deviates from its
mean. Its success has been demonstrated in numerous empirical studies in various
markets like equities~\cite{avellaneda2010statistical},
commodities~\cite{nakajima2019expectations, vaitonis2017statistical}, and
currencies~\cite{fischer2019statistical}; see, \eg,~\cite{gatev2006pairs,
avellaneda2010statistical, perlin2009evaluation, hogan2004testing,
krauss2017deep, caldeira2013selection, huck2019large, dunis2010statistical}. The general setting of a stat-arb is a portfolio of (possibly more than two)
assets that exhibits a mean-reverting behavior~\cite[\S
10.5]{feng2016signal}. The literature on stat-arbs
tends to split into several
categories:
finding stat-arbs,
modeling the (mean-reverting) portfolio price, and trading stat-arbs. For a
detailed overview of the literature, we refer the reader
to~\cite{krauss2017statistical}. 

\paragraph{MBSA.} MBSAs were recently
introduced as an extension of the tradition stat-arb, in which the midpoint of
the price band is allowed to move over time~\cite{johansson2023statarb}.
Although the method is new, it relies on principles related to Bollinger bands that have been known for
decades~\cite{bollinger1992using,bollinger2002bollinger}. The MBSA relies on solving a small convex optimization problem to find a portfolio of
(possibly more than two) assets that vary within a moving band. The method has
been shown to work well in practice, and is used in the empirical study in this
paper.

\paragraph{Trading stat-arbs.}
With some exceptions, the literature on trading stat-arbs is mostly limited to pairs
trading or trading individual stat-arbs, rather than managing a portfolio of
several stat-arbs. Several methods are based on co-integration~\cite{JOHANSEN2000359, alexander2005indexing, granger1983co,
engle_granger_1987, vidyamurthy2004pairs}. For example, in~\cite{zhao2018mean,
zhao2018optimal, zhao2019optimal} the authors consider a (non-convex)
optimization problem for finding high variance, mean-reverting portfolios, in a
co-integration space of several stat-arb spreads. Their strategy is based
on finding a portfolio of
spreads, defined by a co-integration subspace, and implemented using sequential
convex optimization. In \cite{yamada2012model, primbs2018pairs, yamada2018model}
the authors model the spread of a pair of assets as
an autoregressive process, a discretization of the Ornstein-Uhlenbeck
process, and show how to trade a portfolio of spreads under proportional transaction
costs and gross exposure constraints using model predictive control. The authors
of~\cite{yamada2012optimal} use co-integration techniques to find pairs of assets
that are mean-reverting, and show how to construct optimal mean-variance
portfolios of these pairs. 

Our approach differs from the above methods in that we do not
assume any particular model of the price process, and we do not use statistical
analysis like co-integration tests to find stat-arbs. Instead, we assume several
stat-arbs (defined below as a portfolio and a price signal) are given, and the
problem is to find the optimal allocation to these stat-arbs. In other words, we
decouple the problem of finding stat-arbs from the problem of
portfolio construction.

\subsection{Outline} The rest of this paper is structured as follows. In
\S\ref{s-statarb} we review MBSAs and set our notation.
In \S\ref{s-dynamic} we show how to
manage and evaluate a dynamic basket of MBSAs.
We present an empirical study of the method on recent historical data
in \S\ref{s-numerical}, and give conclusions in \S\ref{s-conclusion}.

\section{Moving-band stat-arbs}\label{s-statarb}
MBSAs were recently introduced in~\cite{johansson2023statarb}.
We start with a universe of $n$ assets, with $P_t \in \reals^n_{++}$ denoting
the price (suitably adjusted) in USD, in period $t=1, 2, \ldots$.

\subsection{Midpoint price and alpha}
An MBSA is defined by a
vector $s \in \reals^n$ of asset holdings (in shares),
with negative entries denoting short positions.
The vector $s$ is typically sparse, with only a modest number of nonzero
entries, but that will not affect our method.
The MBSA price in period $t$ is $p_t=s^TP_t$.
The MBSA \emph{midpoint price} is given by the trailing $M$-period average price,
\BEQ \label{eq:midpoint}
\mu_t = \frac{1}{M}\sum_{\tau=t-M+1}^t p_\tau,
\EEQ
where $M$ is the rolling window memory. 
For a good MBSA, the difference of the price and midpoint price,
$p_t-\mu_t$, will oscillate in a band; we obtain a profit by buying 
when the price difference is low and selling when it is high.
We associate with the MBSA the `alpha' value
\BEQ\label{e-alpha}
\alpha_t = \mu_t - p_t.
\EEQ
If $\alpha_t$ is positive, we expect the price of the MBSA to increase, 
and if it is negative, we expect the price to decrease. 

\subsection{MBSA lifetime}
In~\cite{johansson2023statarb}
we show how to discover MBSAs (\ie, the vector $s$) by
approximately solving a constrained variance maximization problem, using
the convex-concave procedure~\cite{shen2016disciplined,
lipp2016variations}. 

Each MBSA has a lifetime, starting at creation or discovery time
$t=d$ and ending at time $t=e>d$, when we choose to decommission it.
We say it is active over the periods $t=d, d+1, \ldots, e$.
(See~\cite[\S2.3]{johansson2023statarb} for details.)

\subsection{Multiple MBSAs} We consider a collection of multiple active MBSAs
that changes over time as new ones are discovered and some existing ones
are decommissioned.  We denote
$K_t$ as the number of MBSAs active during period $t$,
and index the MBSAs as $k=1, \ldots, K_t$.
The $k$th MBSA is defined by the holdings $s^{(k)}\in\reals^n$, which
gives rise to the price $p^{(k)}_t = (s^{(k)})^T P_t$, and alpha
\[
\alpha^{(k)}_t = \mu^{(k)}_t - p^{(k)}_t,    
\]
where $\mu^{(k)}_t$ is the moving midpoint of the $k$th MBSA. 
We collect the holdings of the $K_t$ MBSAs in a matrix 
\[
S_t =[ s^{(1)}\;\cdots \;s^{(K_t)}] 
\in\reals^{n\times K_t},
\]
and introduce the $K_t$-vectors
\[
p_t=(p^{(1)}_t,\ldots,p^{(K_t)}_t), \qquad \mu_t=(\mu^{(1)}_t,\ldots,\mu^{(K_t)}_t), \qquad 
\alpha_t=(\alpha^{(1)}_t,\ldots,\alpha^{(K_t)}_t),
\] 
\ie, we extend the notation
introduced in the previous paragraph to the setting of multiple MBSAs.

\section{Managing a dynamic basket of MBSAs}\label{s-dynamic}
%This section introduces the optimization framework for managing a dynamic basket
%of MBSAs, based on Markowitz mean-variance optimization 
%principles~\cite{markowitz_1952,markowitz_at_seventy}.

\subsection{Portfolio holdings and trades} \label{s-portfolio-holdings}
\paragraph{Arb-level and asset-level portfolio holdings.} The (MBSA) holdings 
during period $t$ are denoted by $q_t \in
\reals^{K_t}$. (The units of $q_t$ are `shares' of the
MBSAs.)  The corresponding asset-level holdings are
$h_t=S_tq_t \in \reals^n$
(in shares), or, in USD,
$P_t \circ (S_t q_t)$,
where $\circ$ denotes the elementwise (Hadamard) product. 

The cash account
is denoted by $c_t \in \reals$ (in USD). 
Since we will use the cash account as part of our collateral for short
positions, we will assume $c_t>0$ for all $t$,
\ie, we do not borrow cash. The total portfolio value at time $t$ is then 
$p_t^T q_t + c_t$ in USD.
 
\paragraph{Trading cost.} We denote the (asset level) 
trades vector at time $t$ as $z_t = h_t- h_{t-1}$, 
\ie, the change in asset-level holdings from time $t-1$ to time $t$, in shares.
The trading cost at time $t$ is given by
\[
    (\kappa_t^{\text{trade}})^T |z_t|,
\]
where $\kappa_t^{\text{trade}} \in \reals^n_+$ is the vector of one half the
bid-ask spread for each asset at time $t$, in USD per share,
and the absolute value is elementwise.

\paragraph{Holding cost.}
The holding cost is given by
\[
(\kappa_t^{\text{short}})^T (-h)_+,
\]
where $\kappa_t^{\text{short}} \in \reals^n$ is the vector of shorting rates for each
asset over period $t$, in USD per share per trading period.
Here $(u)_+=\max\{u,0\}$ is the nonnegative part of $u$, and in
the expression above it is applied elementwise.

\subsection{Markowitz objective and constraints} \label{s-Markowitz}
We manage a dynamic basket of MBSAs using a Markowitz
optimization approach \cite{markowitz_1952, markowitz1959portfolio, boyd2023markowitz},
maximizing the expected return adjusted for transaction and holding costs, 
subject to constraints on the portfolio holdings which include a risk limit.
We assume that we know the current asset prices $P_t$,
the arb-to-asset transformation matrix $S_t$, 
and the previous portfolio holdings $h_{t-1}$ (and $q_{t-1}$) and cash value $c_{t-1}$. 
Our goal is to find the new portfolio holdings $h_t$ (and $q_t$ and $c_t$).
We denote the candidate values, which are optimization variables, as
$h$, $q$, and $c$, respectively.

\paragraph{Objective.} The objective is to maximize the 
alpha exposure, adjusted for transaction and shorting costs,
\[
\alpha_t^T q - \gamma^{\text{trade}} (\kappa_t^{\text{trade}})^T(h-h_{t-1}) 
- \gamma^{\text{short}} (\kappa_t^{\text{short}})^T (-h)_+,
\]
where $\gamma^{\text{trade}}$ and $\gamma^{\text{short}}$ are positive parameters
trading off the alpha exposure, transaction cost, and holding cost. 
The true values of $\kappa_t^{\text{trade}}$ and 
$\kappa_t^{\text{short}}$ are not
known at time $t$, but are estimated from historical data.
This objective is a concave function of the variables $h$ and $q$.

% We soften this constraint by adding a penalty term
% \[
% \gamma^{\text{hold}}\|h - S_tq\|_1
% \]
% to the objective, where $\gamma^{\text{hold}}$ is a positive parameter trading
% off the objective terms. The main reason for softening this constraint is to
% reduce unnecessary trading. After softening, $h=S_tq$ may not hold exactly. We then use the solution $h$ to define the new portfolio.

\paragraph{Cash-neutrality.} We assume that the portfolio is
cash-neutral~\cite{boyd2023markowitz,grinold2000active}, \ie, 
\BEQ \label{eq:cash-neutrality}
    p_t^T q = 0.
\EEQ
In other words, the total portfolio value is equal to the value of the cash account. (We
can also add a limit on the market exposure (or beta), but we found empirically that this makes
a small difference on top of the cash neutrality constraint.) This
constraint is often used in long-short trading strategies~\cite[Chap.
15]{grinold2000active}. 
The cash neutral constraint is a linear equality constraint on the variable $q$.

\paragraph{Collateral constraint.} 
We require that
\[
P_t^T(h)_+ +  (c)_+ \geq \eta   (P_t^T(h)_- + (c)_-), 
\]
where $\eta \geq 1$ is a parameter. 
Here $(u)_- = (-u)_+=\max\{-u, 0\}$ is the nonpositive 
part of $u$, applied elementwise to $h$ above.
This constraint ensures that the long position is at 
least $\eta$ times the short position. 
In the form above, it is not a convex constraint, but
subtracting $(P_t^T(h_t)_- + (c_t)_-)$ from both sides yields the equivalent 
convex constraint 
\[
P_t^Th + c \geq (\eta-1)   (P_t^T(h)_- + (c)_-). 
\]
The cash-neutrality constraint implies $P_t^T h=0$, so this simplifies to
\[
c \geq (\eta-1)  P_t^T(h)_-,    
\]
\ie, the cash account is at least $(\eta-1)$ times the short position. For
example, we can set $\eta=2.02$ to keep a collateral of 102\% of the short
position, as has typically been required by regulation~\cite{d2002market,
geczy2002stocks}.

\paragraph{MBSA size limit constraints.} 
We add the constraints
\[
|q|_k(p_t)_k \leq \xi_t^{(k)} c, \quad k=1,\ldots,K_t,
\]
\ie, the absolute value of the position in the $k$th MBSA is at 
most $\xi_t^{(k)}$ times
the total portfolio value (which is the same as the cash account), 
where $\xi_t^{(k)}$ is a parameter. These constraints serve two purposes.
First, they act as a form of regularization in case there are more MBSAs than
assets, in which case the covariance matrix in the risk constraint
(defined below) will be ill-conditioned or singular.
Second, they limit extreme positions in any single MBSA. During
the active period $t=d, d+1, \ldots, e$ of an MBSA, we set $\xi_t^{(k)}=\xi$ for
$t=d, d+1, \ldots, e-l$, and then linearly decrease $\xi_t^{(k)}$ to zero over
the next $l$ periods, where $l$ is a parameter, say $l=21$ periods, and $\xi$ is
a parameter common to all MBSAs.
In other words we use the size limit constraints to gradually decommission
MBSAs.
These constraints are linear inequality constraints on the variables $q$ 
and $c$, and so are convex.

\paragraph{Risk constraint.}
The traditional definition of risk of a portfolio is the
variance of the portfolio return, expressed as a quadratic form of the
asset weights with an estimated asset return covariance matrix.
Taking the squareroot we obtain the standard deviation of portfolio return,
\ie, its volatility, which we can convert to
USD by multiplying by the portfolio value.

Here we propose a different risk measure, which we have 
found empirically to work better, and is more in line with the
spirit of MBSAs.  Our risk is based on fluctuation of the 
portfolio value around its expected midpoint,
and is directly expressed in USD.
Recall that the portfolio value is $p_t^Tq_t$, which we expect to fluctuate
around the midpoint value $\mu_t^Tq_t$, both of these in USD.
We take the risk to be an estimate of the short term mean square value
of the difference of the portfolio value and the midpoint value,
$(p_t-\mu_t)^T q_t$.
We express this mean-square value (now using $q$, not $q_t$) as
\[
q^T \Sigma_t q ,
\]
where $\Sigma_t$ is an average of the recent values of 
$(p_\tau - \mu_\tau) (p_\tau - \mu_\tau)^T$.
We limit our risk using the constraint
\[
\|\Sigma_t^{1/2} q\|_2 \leq \sigma c,
\]
where $\sigma > 0$ is a parameter setting the target risk 
(which is unitless) and $c$ is the portfolio value.
This risk limit is a convex constraint in the variables $q$ and $c$,
specifically a second-order cone (SOC) constraint \cite[\S 4.4.2]{BoV:04}.

The covariance matrix $\Sigma_t$ can be estimated in many ways; see,
\eg,~\cite{johansson2023covariance}. 
We express it in terms of asset prices as
\[
\Sigma_t = S_t^T \Sigma^{P}_t S_t,
\]
where $\Sigma^{P}_t\in\reals^{n\times n}$ is an estimate of the
short-term covariance matrix of the asset prices.
We estimate $\Sigma^{P}_t$ as follows. First, define the centered price vectors
\[
\tilde P_t = P_t - \bar P_t, \quad t = 1, \ldots, T,
\]
where $\bar P_t$ is the $M$-period rolling window mean of the asset prices.
(This is the same $M$ used to define the MBSA midpoints 
in~\eqref{eq:midpoint}.) Then, $\Sigma^{P}_t$ is an average of the recent values
of $\tilde P_\tau \tilde P_\tau^T$. If a linear
estimator, like a rolling window or exponentially weighted moving average
(EWMA), is used to estimate this expectation, then this is equivalent to
estimating $\Sigma_t$ directly using the same method, \ie, to center the MBSA
prices and then compute an average of the outer product of the centered
prices. We will use the iterated EWMA
predictor~\cite{johansson2023covariance,cov_barrat_2022}, to estimate the
covariance matrix of the centered prices $\tilde P_t$. (This is not equivalent
to estimating $\Sigma_t$ directly with an iterated EWMA, but in practice 
very similar.)

% described in
% appendix~\ref{c-nonneg-means}, to estimate $\Sigma^P_t$. 

\subsection{Markowitz formulation}
We assemble the objective and constraints described above.
We take $h_t$, $q_t$, and $c_t$, as optimal values of 
$h$, $q$, and $c$ in the optimization problem
\BEQ
\begin{array}{ll}
\mbox{maximize} & \alpha_t^T q - \gamma^{\text{trade}}
(\kappa_t^{\text{trade}})^T (h-h_{t-1}) - \gamma^{\text{short}} (\kappa_t^{\text{short}})^T
(-h)_+
% \\ &- \gamma^{\text{hold}}\|h - S_tq\|_1 - \gamma^{\text{neutral}}|p_t^Tq|
\\
\mbox{subject to} &  h = S_tq,  \quad c \geq (\eta-1)P_t^Th_-, \quad c = c_{t-1} + p_t^Tq_{t-1},
% \mbox{subject to} &  c_t\geq (\eta-1)h_-,
\\ &  p_t^Tq = 0, \quad |q|_k(p_t)_k \leq \xi_t^{(k)} c, \quad 
k=1,\ldots,K_t, \\ & \|\Sigma_t^{1/2} q \|_2 \leq \sigma c, 
% \\ &   |q|_k(p_t)_k \leq \xi_t^{(k)} c_t, \quad k=1,\ldots,K_t, \quad\|(\Sigma_t)^{1/2} q \|_2 \leq \sigma c_{t-1}, 

\end{array}
\label{eq:Markowitz}
\EEQ 
with variables $h \in \reals^{n}$, $q \in \reals^{K_t}$, and $c$.
The problem data are:
\BIT
\item $h_{t-1}$, $q_{t-1}$, $c_{t-1}$, the previous portfolio
holdings at the asset-level and arb-level, and cash;
\item $\kappa_t^{\text{trade}}$ and $\kappa_t^{\text{short}}$,
(predictions of) the trading and holding costs;
\item $S_t$, the arb-to-asset transformation matrix;
\item $\eta$, the collateral parameter;
\item $P_t$ and $p_t$, the asset and MBSA prices;
\item $\xi_t^{(k)}$, the MBSA size limits;
\item $\Sigma_t$, a prediction of the short-term covariance matrix of the
MBSA prices;
\item $\sigma$, a target risk, expressed as a fraction of the portfolio value.
\EIT
The problem \eqref{eq:Markowitz} is a convex optimization problem, 
more specifically one that 
can be transformed to a second-order cone program (SOCP).
The solution to this optimization problem gives the new MBSA portfolio
$q_t$, the asset-level portfolio holdings $h_t$, and the cash
account $c_t$ at time $t$.

\paragraph{Extensions and variations.}
We can add additional constraints, such as a leverage constraint,
asset position limits, maximum market exposure, \etc~\cite{boyd2023markowitz}. We
can also soften some constraints, if it is not critical that they are
satisfied exactly and softening improves performance \cite{boyd2023markowitz}.
For example, to soften the arb-to-asset constraint, we remove 
the constraint $h=S_tq$ and add a
penalty term $\gamma^{\text{hold}}\|P_t\circ(h - S_tq)\|_1$ to the objective. 
(Note that softening the arb-to-asset constraint implicitly also softens the risk
and cash-neutrality constraints if these are expressed in terms of the arb-level
portfolio $q$.) 
Softening some constraints
allows the optimizer to choose values that violate the original hard constraints
when necessary, which can reduce unnecessary trading, and therefore transaction cost.

% \paragraph{Softening constraints.}
% A constraint $f=f^\text{des}$ can be softened by adding a penalty term
% $\gamma|f-f^\text{des}|$ to the objective and removing the corresponding
% constraint. This will allow the optimizer to accept solutions that are close to, but not
% exactly, the constraints, and the optimizer will try to avoid the penalty by
% choosing solutions that are close to the constraints. Softening constraints preserves
% convexity of a problem. We will soften the arb-to-asset constraint, and the cash-neutrality constraint,
% by adding a penalty
% \[
% \gamma^{\text{arb-to-asset}}\|h - S_tq\|_1 + \gamma^{\text{neutral}}|p_t^Tq|
% \]
% to the objective in~\eqref{eq:Markowitz}, where $\gamma^{\text{arb-to-asset}}$
% and $\gamma^{\text{neutral}}$ are parameters trading off the objective terms.
% The main reason for softening these constraints is to reduce unnecessary 
% trading. For more details on softening constraints,
% see~\cite[\S4.2]{boyd2023markowitz}. After softening, $h=S_tq$ may not hold
% exactly. We then use the solution $h$ to define the new portfolio. Note that,
% since the risk in~\eqref{eq:Markowitz} is defined in terms of the arb-level
% portfolio $q$, softening the arb-to-asset constraint implicitly also softens the
% risk constraint.

\section{Numerical experiments}\label{s-numerical}
We illustrate the MBSA portfolio management strategy on recent historical data.
Code to replicate the experiments is available at
\begin{quote}
\centering
\url{https://github.com/cvxgrp/cvxstatarb}.
\end{quote}

\subsection{Data and parameters} \label{s-setup}
\paragraph{Data set.}
We gather daily price data from the CRSP US Stock Databases using the Wharton
Research Data Services (WRDS) portal~\cite{WRDS}. The data set consists of
adjusted asset prices of 15405 stocks from January 4, 2010, to 
December 30, 2023, for a total of 3282 trading days.

\paragraph{Monthly search for MBSAs.}
We search for MBSAs every 21 trading days, with the same setup and parameters as described in
detail in~\cite[\S4.1]{johansson2023statarb}. 

\paragraph{Dynamic management of the MBSAs.}
Every day we solve the Markowitz optimization problem~\eqref{eq:Markowitz} to
rebalance our portfolio. Each MBSA is
kept in the portfolio for 500 trading days. After that $\xi$ is reduced linearly
to zero over the next 21 trading days.

\paragraph{Risk model.} As described
in~\S\ref{s-Markowitz}, we decompose the covariance matrix as 
\[
\Sigma_t = S_t^T \Sigma^{P}_t S_t,
\]
where $\Sigma^{P}_t$ is the short-term covariance matrix of the asset prices. The covariance matrix
$\Sigma^{P}_t$ is estimated using the iterated
EWMA (IEWMA) predictor (see~\cite{DCC} 
and~\cite[Ch.~2.5]{johansson2023covariance}) on the centered 
prices $\tilde P_t = P_t - \bar P_t$,
where $\bar P_t$ is the 21-day rolling window mean of the asset prices. For the
IEWMA predictor, we use a 125-day half-life for volatility estimation, and a 
250-day half-life for correlation estimation. 
% for the case of time series with zero mean. We explain the extension to nonzero
% means in appendix~\ref{c-nonneg-means}. We use the half-lives listed in
% table~\ref{t-half-lives}.
% \begin{table}
% \centering
% \begin{tabular}{ll}
% \toprule
%  {Half-life} & {Description} \\
% \midrule
% 21 days & $H^{\text{m}_1}$: half-life of mean estimation in first iteration \\
% 250 days & $H^{\text{vol}}$: half-life of volatility estimation (in first iteration) \\
% 63 days & $H^{\text{m}_2}$: half-life of mean estimation in second iteration \\
% 500 days & $H^{\text{cor}}$: half-life of correlation estimation (in second
% iteration) \\
% \bottomrule
% \end{tabular}
% \caption{Half-lives used in the IEWMA predictor.}
% \label{t-half-lives}
% \end{table}
To reduce trading
induced by the risk-model,
we smooth the
covariances with a 250-day half-life EWMA~\cite[Chap. 9.2]{johansson2023covariance}.

\paragraph{Parameters.}
For the MBSA problem we use the same parameters as
in~\cite[\S4.1]{johansson2023statarb}. For the Markowitz optimization problem we
use $\gamma^{\text{trade}}=1$, $\eta=1$, $\xi=1$, and $\sigma^{\text{tar}}=10\%$
annualized. 

\paragraph{Trading and shorting costs.} Our numerical experiments take into account 
transaction costs, \ie, we buy assets at the ask price, 
which is the (midpoint) price
plus one-half the bid-ask spread, and we sell assets at the bid price,
which is the price minus one-half the bid-ask spread. We use 0.5\% as a proxy for
the annual shorting cost of stocks, which is well above what is typically
observed in practice for liquid stocks~\cite{d2002market, geczy2002stocks,
kim2023shorting, drechsler2014shorting}. We also note that we have tested the
method with shorting costs upwards of 10\% annually, and it remains profitable.

\subsection{Simulation and metrics}
This section explains how we simulate the portfolio, and the metrics we use to
evaluate the performance.

\paragraph{Cash account.}
We initialize the portfolio with a cash account at time $t=0$,
$c_0 = 1$.
The cash account evolves as
\[
c_{t+1} = c_t - (q_{t+1}-q_t)^Tp_{t+1} - \phi_t = c_t + q_t^Tp_{t+1} - \phi_t, \quad t=0,\ldots, T,
\]
where $\phi_t$ is the transaction and holding
cost, consisting of the trading cost at time $t+1$ and the holding cost
over period $t$. (The last equality follows from the cash-neutrality constraint.)

\paragraph{Portfolio NAV.}
The net asset value (NAV) of the portfolio, including the cash
account, at time $t$ is 
\[
V_{t} = c_t + q^T_t p_t = c_t.
\]
Due to the market neutrality constraint in
\eqref{eq:Markowitz} we have $V_t=c_t$, \ie, the NAV is equal to the cash account. 

\paragraph{Return.}
The return at time $t$ is 
\[
r_t = \frac{V_t - V_{t-1}}{V_{t-1}}, \quad t = 1, \ldots, T.
\]
We report several
standard metrics based on the returns $r_t$. The average return is 
\[
\overline r = \frac{1}{T} {\sum_{t=1}^{T}{r_t}},
\]
which we multiply by $250$ to annualize. 
% (Here $T=3282$ is the total number of
% trading days.)
The return volatility is 
\[
\left(\frac{1}{T} \sum_{t=1}^{T}\left(r_t-\overline r\right)^2\right)^{1/2},
\]
which we multiply by $\sqrt{250}$ to annualize.
The annualized Sharpe ratio is the ratio of the annualized average return to the
annualized volatility. Finally, the maximum drawdown is
\[
    \max_{1 \leq t_1 < t_2\leq T} \left(1- \frac{V_{t_2}}{V_{t_1}}\right),
\]
the maximum fractional drop in value form a previous high.

\paragraph{Turnover.}
The turnover at time $t$ is
\[
T = \frac{1}{2} \sum_{i=1}^{n} |((P_t\circ h_t)_i - (P_t\circ h_{t-1})_i)/V_t| = \frac{1}{2} \|P_t\circ (h_t - h_{t-1})\|_1 / V_t,
\]
which we multiply by $250$ to annualize. It measures the amount of trading in the portfolio~\cite[Chap.~16]{grinold2000active}. For example, a turnover of $0.01$ means that the average of total amount bought and
total amount sold is 1\% of the total portfolio value.

\paragraph{Active return and risk.} We define the active return as the return
of the portfolio minus the return of the market, which we take to be the S\&P 500. The active risk is the standard
deviation of the active return. 

\paragraph{Residual return and risk.} Given the portfolio returns
$r_1,r_2,\ldots, r_T$, and the corresponding market returns $r^m_1,r^m_2,\ldots,
r^m_T$, we construct the linear model 
\[
r_t = \beta r^m_t + \theta_t, \quad t=1,\ldots,T,
\]
where $\beta r^m_t$ is the return explained by the market (with $\beta\in\reals^n$), and $\theta_t\in\reals^n$ is the
residual return at time $t$. The residual risk is the standard deviation of the
residual return. The mean residual return is often referred to as the alpha of
the portfolio~\cite{grinold2000active} (not to be confused with the alpha of an MBSA). The information ratio is the ratio of
the portfolio alpha to the residual risk.

\subsection{Results}
\paragraph{Portfolio performance.}
The portfolio performance is summarized in table~\ref{t-portfolio_metrics}.
\begin{table}
\centering
\begin{tabular}{ccccc}
\toprule
 {Return} & {Volatility} & {Sharpe ratio} & {Turnover}& {Drawdown} \\
\midrule
 19\% & 12\% & 1.61 & 136 & 15\% \\
\bottomrule
\end{tabular}
\caption{Metrics for portfolio of MBSAs.}
\label{t-portfolio_metrics}
\end{table}
We attain an average annual return of 19\% at an annual volatility of 12\%,
corresponding to a Sharpe ratio of 1.61, with a
maximum drawdown of 15\% over the roughly 10-year period.
The average turnover is 136, which corresponds to a daily turnover
of roughly 50\%. In comparison to other stat-arb strategies in the
literature, this seems like a reasonable level of turnover~\cite{guijarro2021deep}.

\paragraph{Active and residual return and risk.} 
Here we compare the MBSA strategy to the market, represented by the S\&P 500
with an initial investment of \$1 and diluted with cash to attain the same
annualized risk as the MBSA strategy.
Table~\ref{t-market-comparison} gives a numerical summary. The MBSA strategy
attains an annualized Sharpe ratio of 1.61 (compared to 0.66 for the market)
with a residual return (alpha) of 18\% and a market beta of 11\%. The information ratio is 1.53.
\begin{table}
\centering
\begin{tabular}{cccccc}
\toprule
{Active return} & {Active risk} &  {Residual return} & {Residual risk} & {Beta} & {Information ratio} \\
\midrule
 8\% & 20\% & 18\% & 11\% & 11\% & 1.53 \\
\bottomrule
\end{tabular}
\caption{Comparison of portfolio of MBSAs to the market.}
\label{t-market-comparison}
\end{table}

\begin{figure}
\centering
% \hspace{6.9em}
\subfigure[NAV evolution.]{\includegraphics[width=0.75\textwidth]{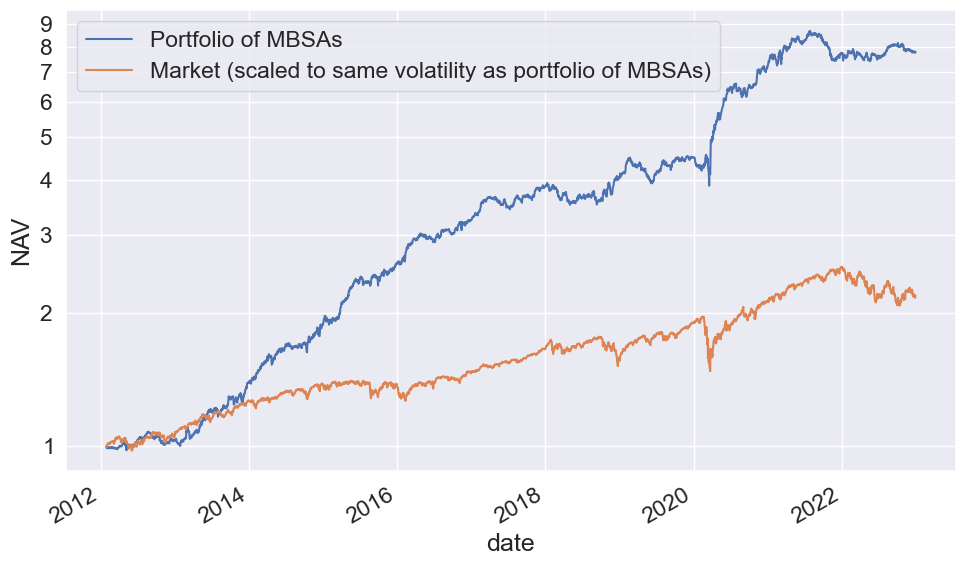}}
\\\subfigure[250-day half-life EWMA correlation to market.]{\includegraphics[width=0.75\textwidth]{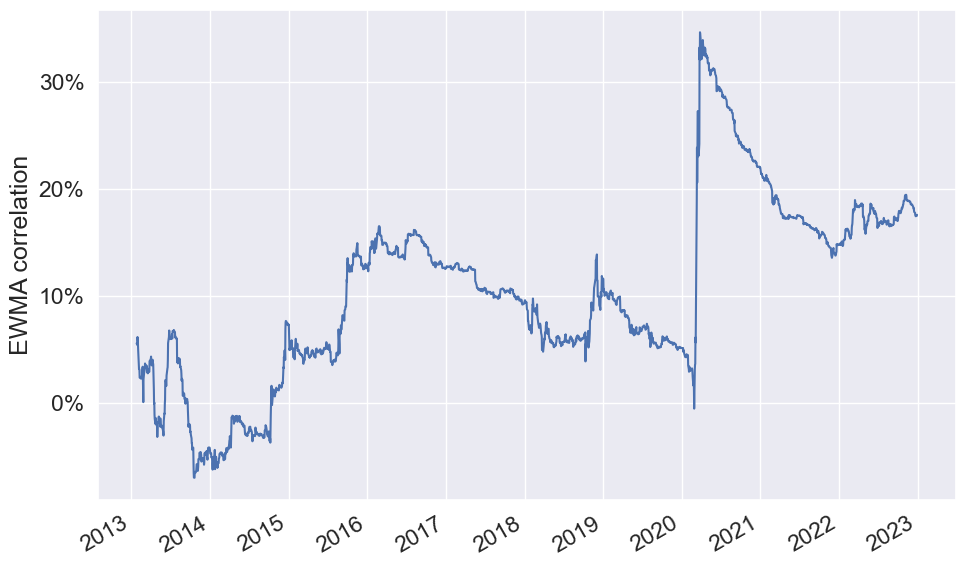}}
\caption{A comparison summary of the dynamically managed portfolio of MBSAs to the market.}
\label{f-market-comparison}
\end{figure}

\paragraph{NAV evolution.}
Figure~\ref{f-market-comparison} shows the NAV of the MBSA strategy and the
market, and the 250-day half-life EWMA correlation between the two.
As seen, the MBSA strategy outperforms the market, with very low
correlation, 15\% over the whole period. 
This suggests mixing the MBSA strategy with the
market. For example, mixing 90\% of the MBSA strategy
and 10\% of the market yields an annualized Sharpe ratio of 1.66, slightly
better than the MBSA strategy alone.

\begin{figure}
\centering
\subfigure[Annual returns.]{\includegraphics[width=0.6\textwidth]{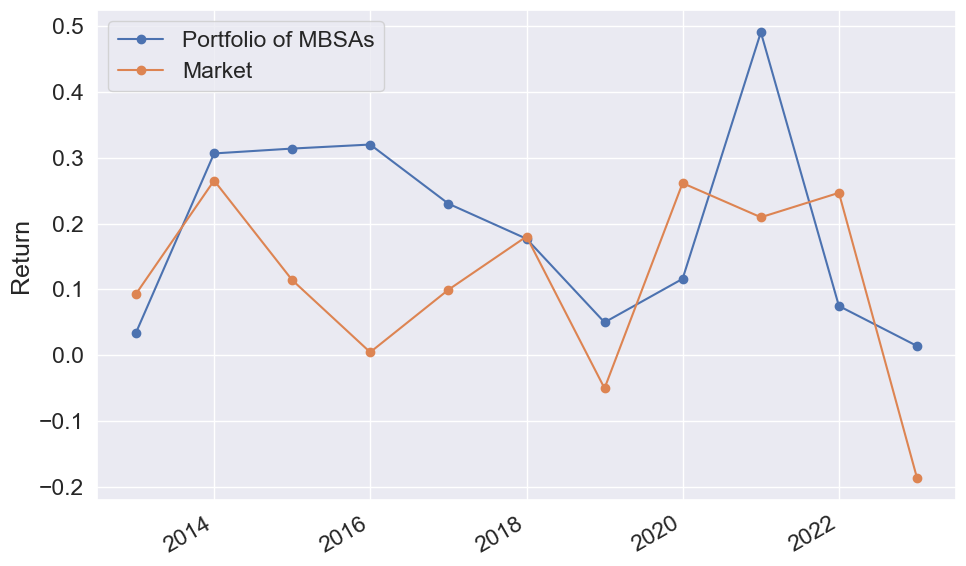}}
\\\subfigure[Annual volatilities.]{\includegraphics[width=0.6\textwidth]{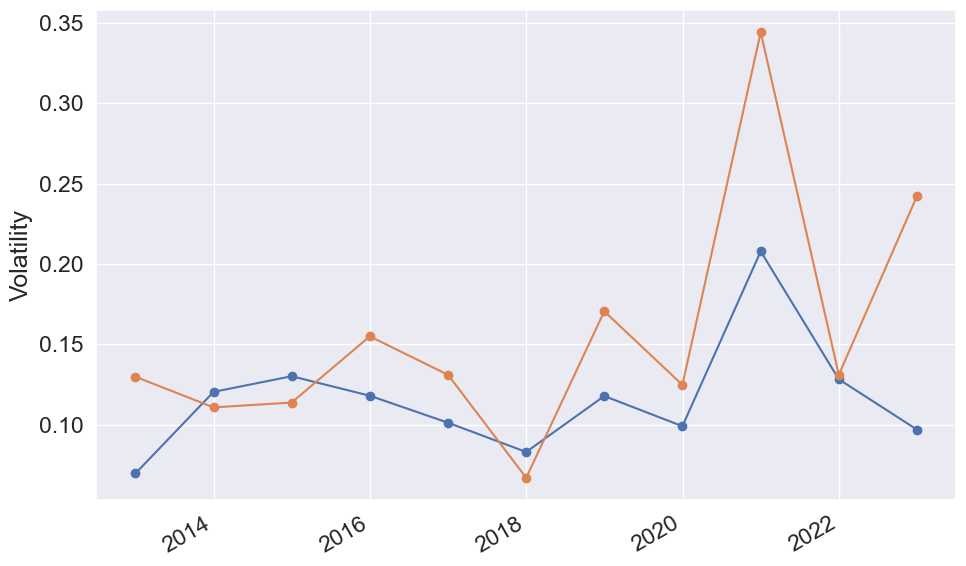}}
\\\subfigure[Annual Sharpe ratios.]{\includegraphics[width=0.6\textwidth]{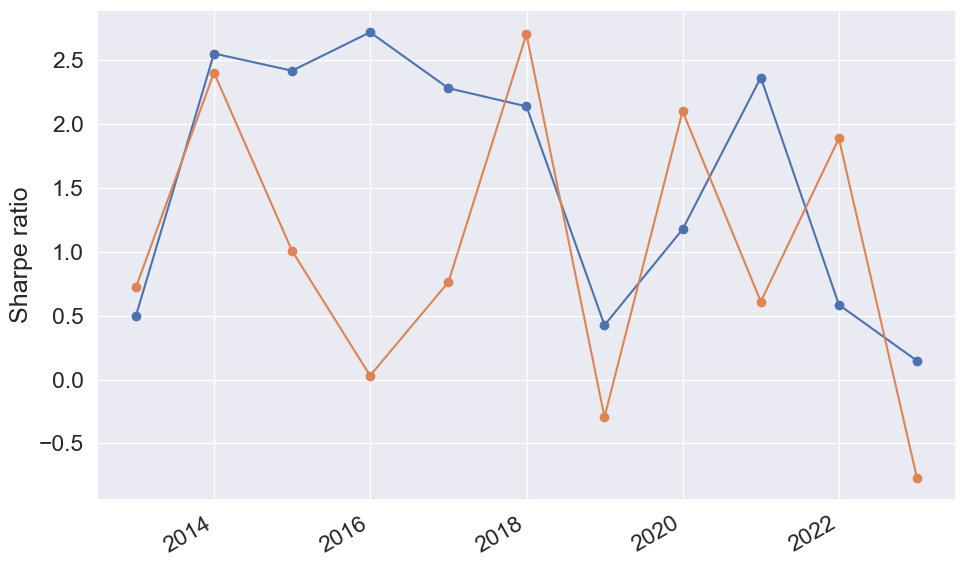}}
\caption{Annual portfolio performance.}
\label{f-annual}
\end{figure}

\clearpage
\paragraph{Annual performance.}
Here we break down the metrics reported above over the 11-year period into
the performance metrics for each year.
%Figure~\ref{f-turnover} shows the annual turnover over time of the MBSA strategy.
%The turnover initially grows to about 200, and then decreases and oscillates
%around 100. 
%\begin{figure}
%\centering
%\includegraphics[width=0.75\textwidth]{figures/portfolio/turnover.png}
%\caption{Annualized turnovers of the MBSA strategy.}
%\label{f-turnover}
%\end{figure}
Figure~\ref{f-annual} shows the annual performance of the MBSA strategy and the
market. The MBSA strategy has positive return in each of the 11 years, whereas
the market has negative return in 2 of the 11 years.
The MBSA strategy outperforms the market in 8 of the 11 years, and has
more stable performance.
% three vertical figures

\begin{figure}
\centering
\subfigure[Annual residual returns (alphas).]{\includegraphics[width=0.6\textwidth]{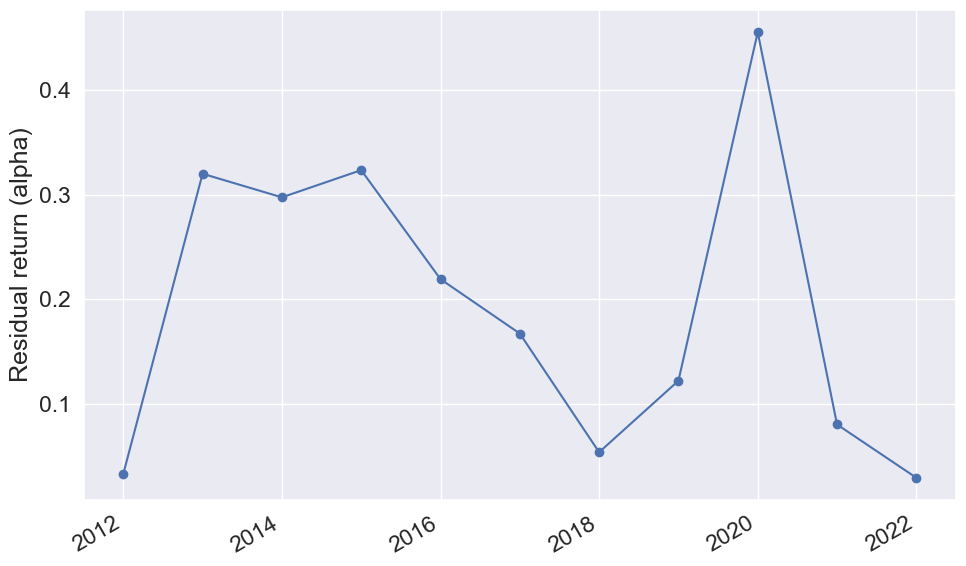}}
\\\subfigure[Annual residual risks.]{\includegraphics[width=0.6\textwidth]{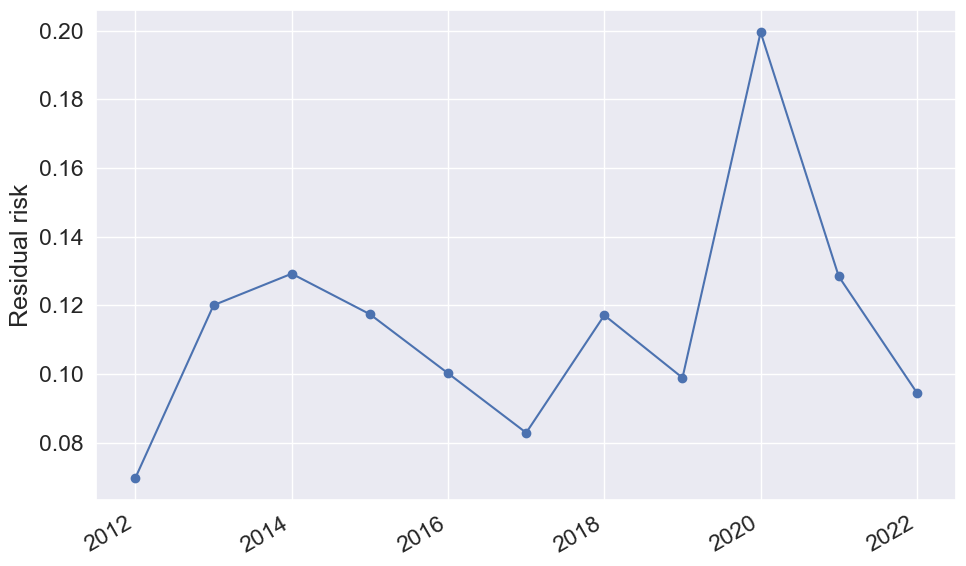}}
\\\subfigure[Annual market betas.]{\includegraphics[width=0.6\textwidth]{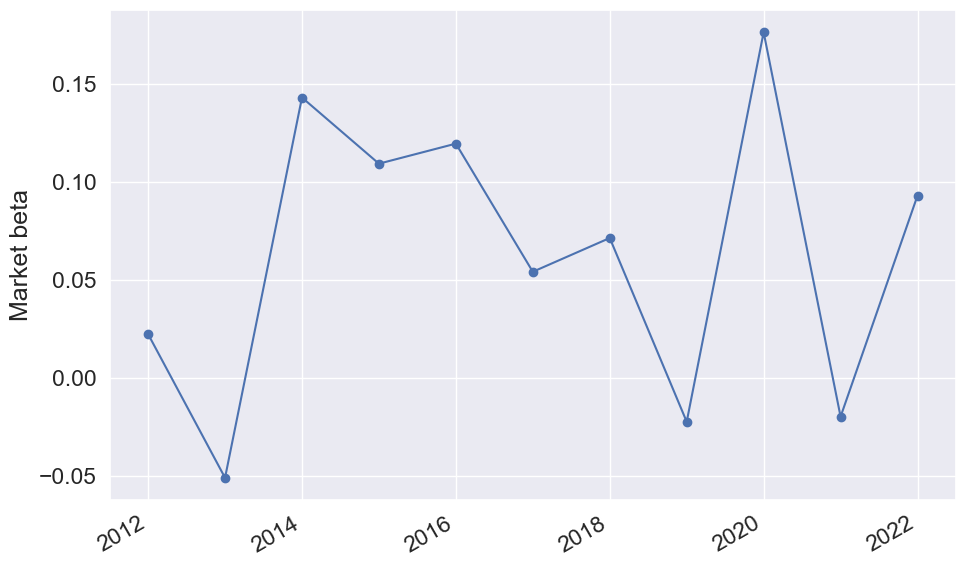}}
\caption{Annual residual return and risk, and market beta.}
\label{f-annual-residual}
\end{figure}

Finally, figure~\ref{f-annual-residual} shows the annual residual return and
risk, and market beta. As seen, most of the MBSA success is unexplained by the
market.

\clearpage
\section{Conclusion}\label{s-conclusion}
We have shown how to manage a dynamic basket of moving-band stat-arbs, based on
a long-short Markowitz optimization strategy. We presented an empirical study of the method on recent
historical data, showing that it can to outperform the market
with low correlation. 

% \paragraph{Constraints.}
% Common constraints encoded in $\mathcal W$ include a maximum leverage constraint
% $\|w\|_1 \leq L_{\mathrm{max}}$, asset position limits $w_{\mathrm{min}} \leq w
% \leq w_{\mathrm{max}}$, and .

% \subsection{Interpretation via a simple trading policy}
% For $K_t=1$, \ie, a single stat-arb, the Markowitz optimization problem (without
% the additional constraints $q\in\mathcal{Q}$ and $h\in\mathcal{H}$) simplifies
% to
% \[
% \mbox{maximize} \quad \alpha_t q - \gamma \sigma_t q^2,    
% \]
% where $\sigma_t^2=s^T \Sigma_t s$ and $s$ is the vector of the stat-arb's
% holdings (in units shares). The solution is 
% \[
% q = \frac{\alpha_t}{\gamma \sigma_t} = \frac{\mu_t-p_t}{\gamma \sigma_t},
% \]
% \ie, it simplifies to 

\bibliography{refs}

\end{document}